\documentclass[aps,prb,twocolumn,showpacs]{revtex4-1}

\usepackage[latin1]{inputenc}
\usepackage[english]{babel}
\usepackage{setspace}  
\usepackage{graphicx}
\usepackage{amssymb}
\usepackage{amsmath}
\usepackage{amscd}
\usepackage{rotating}
\usepackage{color}
\usepackage{epstopdf}
\usepackage{enumerate}
\usepackage{hyperref}

\usepackage{amsthm}

\long\def\beginpgfgraphicnamed#1#2\endpgfgraphicnamed{\includegraphics{#1}}

\begin{document}

\newcommand{\zz}{\mathbb{Z}_2}
\newcommand{\zq}{\mathbb{Z}_q}
\newcommand {\uo}{U(1)}
\newcommand{\ord}[1]{\mathcal{O}(#1)}
\newcommand{\Hs}{\mathcal{H}}

\newcommand{\BIGOP}[1]{\mathop{\mathchoice%
{\raise-0.22em\hbox{\huge $#1$}}%
{\raise-0.05em\hbox{\Large $#1$}}{\hbox{\large $#1$}}{#1}}}
\newcommand{\bigtimes}{\BIGOP{\times}}

\title{Implementing global Abelian symmetries in projected entangled-pair state algorithms}

\author{B. Bauer$^1$, P. Corboz$^{1,2,3}$, R. Or\'us$^{3}$, M. Troyer$^1$}
\affiliation{
$^1$Theoretische Physik, ETH Zurich, 8093 Zurich, Switzerland \\
$^2$Institute of Theoretical Physics, École Polytechnique Fédérale de Lausanne, CH-1015 Lausanne, Switzerland \\
$^3$School of Mathematics and Physics, The University of Queensland, QLD 4072, Australia
}

\date{\today}

\begin{abstract}
Due to the unfavorable scaling of tensor network methods with the refinement parameter $M$, new approaches are necessary to improve the efficiency of numerical simulations based on such states in particular for gapless, strongly entangled systems. In one-dimensional DMRG, the use of Abelian symmetries has lead to large computational gain. In higher-dimensional tensor networks, this is associated with significant technical efforts and additional approximations. We explain a formalism to implement such symmetries in two-dimensional tensor network states and present benchmark results that confirm the validity of these approximations in the context of projected entangled-pair state algorithms.
\end{abstract}



\maketitle

\section{Introduction}

The density matrix renormalization group (DMRG)~\cite{white1992} and matrix-product states (MPS)~\cite{ostlund1995} have proven to be extremely powerful algorithms for one-dimensional quantum systems. For higher-dimensional systems, however, they scale unfavorable with the system size. The reason for this is found in the scaling of entanglement entropy, which is for many systems governed by the area law. This scaling cannot be correctly captured with MPS.

Other ansatz states have been proposed that by construction obey the correct scaling of the entanglement entropy. Prominent classes of such states are \emph{projected entangled-pair states} (PEPS) 
\cite{verstraete2004,sierra1998,nishino1998,nishino2000,nishio2004,murg2007,
pchen2009,nishino2001,maeshima2001,gendiar2003,gendiar2005,isacsson2006,jiang2008,gu2008,murg2009,xie2009,jordan2009,orus2009,orus2009-1,bauer2009,pchen2010}
and the \emph{multi-scale entanglement renormalization ansatz} (MERA).\cite{vidal2007-1,vidal2008,evenbly2009-1,evenbly2009-2,evenbly2010,pfeifer2009} Just as DMRG/MPS, these ansatz states cover the full Hilbert space of the quantum problem with a systematic refinement parameter $M$. In MPS algorithms, the scaling of the computational complexity with this refinement parameter is $\ord{M^3}$, where the number of variational parameters grows as $\ord{M^2}$. In the case of, e.g., PEPS on an infinite square lattice, the scaling of computational complexity is $\ord{M^{12}}$ while the number of variational parameters grows as $\ord{M^4}$. This extremely fast increase of computational effort severely limits the attainable $M$ to currently about $M=2 \ldots 8$.

Previous work \cite{bauer2009} has shown that the accuracy that can be obtained with such limited bond dimensions is very limited in particular for gapless with a large symmetry group. In order to make progress towards controversial problems in condensed matter theory, it is therefore necessary to significantly improve the accuracy of tensor-network state methods by reaching larger bond dimensions $M$.

In one-dimensional DMRG calculations, exploiting global Abelian symmetries has led to large improvements of the accuracy.\cite{schollwoeck2005} Non-Abelian symmetries have also been considered.\cite{sierra1997,dukelsky1998,wada2000,wada2001,mcculloch2000,mcculloch2001,mcculloch2002} In the context of two-dimensional tensor network state calculations, symmetries have only been explored very recently. Parity symmetry ($\zz$) plays a central role in the definition of fermionic tensor networks\cite{kraus2010,barthel2009,corboz2009,corboz2009-1,corboz2010,shi2009,pizorn2010,gu2010} but has also been shown to be useful for spin systems.\cite{cincio2008} Continuous groups, such as $\uo$, have been used in calculations with the TERG algorithm \cite{zhao2010} and the MERA.\cite{evenbly2010-1,singh2010} A general introduction to the topic without numerical results is given in Ref.~\onlinecite{singh2010-1}; Ref.~\onlinecite{singh2010} contains a detailed introduction to $\uo$ symmetry and its use for MERA computations.

In this paper, we will develop a formalism to implement Abelian symmetries into tensor network states. We will study the example of infinite projected entangled-pair states and numerically confirm the validity of the approximation introduced by restricting the structure of the tensors.

\subsection{Projected entangled-pair states}
\begin{figure}
\centering
\beginpgfgraphicnamed{fig_peps}
\begin{tikzpicture}
\foreach \i in {1,...,3}
\foreach \j in {1,...,3}
{ {
        \node (G\i\j) at (1.5*\i+1*\j,1.5*\j) [inv] {};
} }

\foreach \k / \i in {1/2}
\foreach \l / \j in {1/2}
{ {
        \node at (G\i\j) [peps] {};
        \draw[thick] (G\i\j) -- node[below right ] {$d$} +(0,-1);
        \node at ($ (G\i\j)+(0,-1.3) $) {$\phi_i$};
} }
\foreach \i in {2}
\foreach \j / \l in {1/2,2/3}
{ {
        \draw[thick] (G\i\j) -- node[left] {$M$} (G\i\l);
        \draw[thick] (G\j\i) -- node[above] {$M$} (G\l\i);
} }

\foreach \i in {1,...,4}
\foreach \j in {1,...,4}
{ {
        \node (G\i\j) at (2.5+1.2*\i+0.8*\j,1*\j) [inv] {};
} }

\foreach \i / \j / \k in {2/2/1,2/3/2,2/4/3,3/2/4,3/3/5,3/4/6,4/2/7,4/3/8,4/4/9}
{
        \node at (G\i\j) [peps] {};
        \draw[thick] (G\i\j) -- node[below right ] {$\phi_{\k}$} +(0,-0.6);
}
\foreach \i in {2,3,4}
\foreach \j / \l in {2/3,3/4}
{ {
        \draw[thick] (G\i\j) -- (G\i\l);
        \draw[thick] (G\j\i) -- (G\l\i);
} }
\end{tikzpicture}
\endpgfgraphicnamed
\caption{Pictorial representation of a projected entangled-pair state (PEPS). Left panel: For the square lattice, a tensor of rank 5 will be associated with each lattice site. The index pointing down connects to the physical system, while the other indices connect to neighboring tensors in the state. Right panel: The panel shows the PEPS decomposition of a coefficient $c(\phi)$ for a state $|\Psi\rangle = \sum c(\phi_1 \ldots \phi_9) |\phi_1 \ldots \phi_9\rangle$ on a $3 \times 3$ square lattice with open boundary conditions. \label{fig:peps}}
\end{figure}
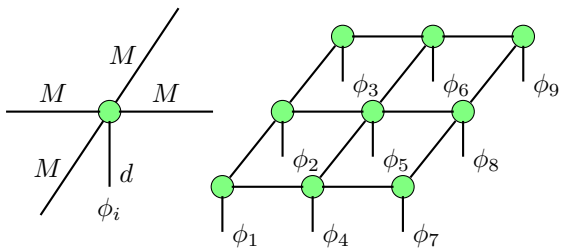

Let us now turn to a short introduction of projected entangled-pair states. We consider a lattice system with a tensor-product Hilbert space $\Hs = \bigotimes \Hs_i$ and a product basis $\lbrace |\phi\rangle = |\phi_1\rangle |\phi_2\rangle \ldots \rbrace$. In order to approximate the coefficients $c(\phi)$ of a wave function $|\Psi\rangle = \sum c(\phi) |\phi\rangle$, we associate with each site of the physical lattice a tensor of rank $z+1$, where $z$ is the number of nearest neighbors of the site. In this paper, we will focus on the square lattice, where $z=4$. A graphical representation is shown in Figure~\ref{fig:peps}. Of these $z+1$ indices, one is considered the physical index of the tensor with dimension $d = \text{dim} \Hs_i$, whereas the other ones are auxiliary indices connecting to the nearest neighbors with dimension $M$. The coefficient $c(\phi)$ is then given as the trace over all auxiliary indices in the network.

To represent a lattice with $N$ inequivalent sites, usually $N$ different tensors have to be optimized. We can however assume that the system is invariant under translations by a certain number of sites. Such a state can be represented with only few independent tensors and the thermodynamic limit can be taken directly.

PEPS are a higher-dimensional generalization of matrix-product states. They inherit important properties from MPS: i) For $M=1$, they are equivalent to static mean-field theory. ii) They can capture the entanglement properties of gapped systems in the sense that the rank of reduced matrices for a block of sites is bounded by the exponential of the surface of the block, which allows the entanglement entropy to diverge with an area law. Unlike matrix-product states, however, the exact evaluation of expectation values can in general not be performed in polynomial time. Therefore, approximate methods are required. Several such methods have been proposed.\cite{verstraete2004,jordan2008,jiang2008,orus2009-1} They all have in common that they lead to polynomial scaling, yet with large exponents.

\section{Symmetry groups}

\subsection{Charge calculus} \label{sct:nomenclature}

In the following, we will be concerned with Hamiltonians $H$ with a symmetry group $\mathcal{G}$, i.e. they commute with the elements of some Abelian group $\mathcal{G}$, $[H, q] = 0\ \forall q \in \mathcal{G}$. This implies that eigenstates of $H$ are also eigenstates of $q$. We require:
\begin{itemize}
\item There exists a unitary representation $U$ of the group. For $q \in \mathcal{G}$, we have $U^T(q^{-1}) = U(q)$.
\item All representations of the group decompose into a direct sum of irreducible representations $\mathcal{V}_i$, which can be labelled in correspondence to the eigenvalues of some operator $g$. We will call these labels $c_i$.
\item Consider a state $|\phi\rangle \in \mathcal{V}_i$ and $q \in \mathcal{G}$. We then have
\begin{equation} \label{eqn:nu}
U(q) |\phi\rangle = \nu(q, c_i) |\phi\rangle
\end{equation}
with $\nu(q, c_i) \in \mathbb{C}$.
\end{itemize}
Examples will be discussed in Section~\ref{sct:groups}.

We can classify eigenstates of the Hamiltonian $H$ into the irreducible representations of $\mathcal{G}$. The associated labels $c_i$ are then called good quantum numbers (for brevity, we will also call them charges or symmetry sectors).

As we will see below, only few properties of the irreducible representations of the group are needed to implement symmetric tensor networks. These are intimately related to the properties of the eigenvalues defined in Eq.~\eqref{eqn:nu}.
\begin{description}

\item[Charge fusion] Consider two states $|a\rangle \in \Hs_1, |b\rangle \in \Hs_2$ with associated quantum number $c_1$ and $c_2$, respectively. Their tensor product in $\Hs_1 \otimes \Hs_2$ also has a well-defined quantum number $c_3$. We thus define the fusion of quantum numbers
\begin{equation}
c_1 \times c_2 = c_3.
\end{equation}
This corresponds to a labeling of the tensor product of irreducible representations. For the eigenvalues $\nu$, this corresponds to
\begin{equation} \label{eqn:numult}
\nu(q, c_1) \nu(q, c_2) = \nu(q, c_3).
\end{equation}

\item[Identity charge] There exists an identity charge $\mathbb{I}$ such that $c \times \mathbb{I} = c\ \forall c$. This implies $\nu(q, \mathbb{I}) = 1$.

\item[Conjugate charge] For each charge $c$, a conjugate charge $\bar{c}$ exists such that
\begin{equation}
c \times \bar{c} = \mathbb{I}
\end{equation}
This imples $\nu(q, \bar{c}) = 1/\nu(q, c)$.

\end{description}

We can easily generalize the above to products of groups. For $\tilde{\mathcal{G}} = \mathcal{G}^1 \times \mathcal{G}^2$, the irreducible representations are $\tilde{\mathcal{V}}_{ij} = \mathcal{V}^1_i \times \mathcal{V}^2_j$, which can be labelled by $\tilde{c}_{ij} = (c_i^1, c_j^2)$. These labels correspond to eigenvalues of the operator $\tilde{g} = (g^1 \otimes \mathbb{I}, \mathbb{I} \otimes g^2)$. The above calculus is then constructed from element-wise operations on the $\tilde{c}$.

\subsection{Examples} \label{sct:groups}

An important example is the $\uo$ symmetry, which is present in systems with particle number conservation and many spin models. For benchmarking purposes, we will apply the symmetric PEPS algorithm to a system of spin-$\frac{1}{2}$ degrees of freedom on the square lattice with Heisenberg interaction. This system has an $SU(2)$ spin rotation symmetry, which in the thermodynamic limit and at zero temperature is spontaneously broken to a $\uo$ symmetry. We will exploit this group and its finite subgroups.

\subsubsection{$\zz$}
For Hamiltonians that are invariant under a simultaneous flip of all spins, $|\uparrow\rangle \leftrightarrow |\downarrow\rangle$, the operator
\begin{equation}
g_{\zz} = (-1)^{\sum_i \sigma_i^z} = \prod_i \sigma_i^z
\end{equation}
commutes with the Hamiltonian. A unitary representation of the group $\zz$ is given by
\begin{equation}
U(\alpha) = g_{\zz}^\alpha
\end{equation}
with $\alpha \in \lbrace 0,1 \rbrace$. Its unitarity follows from the unitarity of the Pauli matrix $\sigma^z$. The two irreducible representations can be labeled as $c = \pm $. The eigenvalues $\nu(\alpha, c)$ are:
\begin{eqnarray}
\nu(0, +) = +1 &\ \ \ \ \ &\nu(1,+) = +1 \\
\nu(0, -) = +1 &\ \ \ \ \ &\nu(1,-) = -1
\end{eqnarray}
The fusion rules therefore are:
\begin{equation}
\pm \times \pm = + \ \ \ \ \pm \times \mp = - \\
\end{equation}
This implies  $+ = \mathbb{I}$ and $\bar{c} = c$.

Due to the very simple structure with only two irreducible representations, the implementation of $\zz$ symmetry is particularly easy.

\subsubsection{$\uo$} \label{sct:u1}
The most commonly used symmetry in simulations with exact diagonalization or DMRG is the $U(1)$ spin symmetry, which is given if the operator
\begin{equation}
g_{\uo} = \sum_i \sigma_i^z
\end{equation}
commutes with the Hamiltonian. This is the infinitesimal generator of a representation of $U(1)$,
\begin{equation}
U(\phi) = \exp \left( i2\pi \phi g_{\uo} \right),
\end{equation}
where $\phi \in [0,2\pi)$.

The irreducible representations can be labeled with integer numbers, $c \in \mathbb{Z}$. The $\nu(\phi, c)$ are
\begin{equation}
\nu(\phi, c) = \exp \left( i2\pi \phi c \right).
\end{equation}
Clearly,
\begin{eqnarray}
\nu(\phi, c_1) \nu(\phi, c_2) &=& \nu(\phi, c_1 + c_2) \\
\nu(\phi, 0) &=& 1 \\
\nu(\phi, -c) &=& \nu(\phi, c)^{-1}
\end{eqnarray}
The charge calculus therefore follows the rules of integer addition. The label of the irreducible representations can be interpreted as magnetization of the state. Special care must be taken when forming the adjoint of a vector or operator, since $U(\phi) |\Psi\rangle \rightarrow \langle\Psi | U(-\phi)$. The Hermitian transpose of a state in the irreducible representation $c$ therefore falls into the irreducible representation $\bar{c}$.

\subsubsection{$\zq$}
Since for the PEPS, finite groups are easier to deal with, we consider finite subgroups of $\uo$, namely the cyclic groups $\zq$. We define
\begin{equation}
g_{\zq} = \exp \left( \frac{i 2 \pi}{q} \sum_i \sigma_i^z \right),
\end{equation}
which naturally also commutes with the Hamiltonian if $g_{\uo}$ does. The irreducible representations can be labeled with $c \in \lbrace 0, \ldots, q \rbrace$, where 0 is the identity. A unitary representation is, similar to $\zz$, given by
\begin{equation}
U(\alpha) = g_{\zq}^\alpha.
\end{equation}
where $\alpha \in \lbrace 0, \ldots, q-1 \rbrace$. The eigenvalues $\nu(\alpha, c)$ are
\begin{equation}
\nu(\alpha, c) = \left( e^{i\frac{2\pi}{q}} \right) ^{\alpha c}.
\end{equation}
This implies the cyclic property $\nu(\alpha, c+q) = \nu(\alpha+q, c) = \nu(\alpha, c)$. The resulting fusion rule is
\begin{equation}
c_1 \times c_2 = (c_1 + c_2) \mod q,
\end{equation}
therefore
\begin{equation}
\bar{c} = q - c.
\end{equation}
For taking adjoints, the same consideration as in the case of $\uo$ applies.
The implementation of $\zq$ symmetry for $q > 2$ is more involved than $\zz$ since charges are not inverse to themselves. The small number of sectors however reduces the technical efforts.

\section{Symmetric tensor networks}

\subsection{Definition and contraction of symmetric tensors} \label{sct:symmtensor}

We define a tensor $T$ as a linear map from a tensor product of Hilbert spaces to the complex numbers:
\begin{equation} \label{eqn:tensor}
T: \Hs_1 \otimes \Hs_2 \otimes \ldots \otimes \Hs_R \rightarrow \mathbb{C}.
\end{equation}
Here, $R$ is the rank of the tensor. The elements of the tensor are $T(v_1, v_2, \ldots)$ for $v_k \in \Hs_k$. Equivalently, if we choose a fixed basis $\lbrace b_i^k \rbrace$ in each $\Hs_k$, we can define a tensor as a multidimensional array $T_{i_1 i_2 i_3 \ldots}$, where the indices $i_k$ run from 1 to $\dim \Hs_k$ and
\begin{equation}
T_{i_1 i_2 i_3 \ldots} = T(b^1_{i_1}, b^2_{i_2}, b^3_{i_3}, \ldots).
\end{equation}

In this paper, we are interested in states composed of tensors that are invariant under the operations of a group. To define this, let $q \in \mathcal{G}$ and $U^k(q)$ unitary representations in the Hilbert spaces $\Hs_k$. We then require
\begin{equation} \label{eqn:inv}
T(U^1(q) v_1, U^2(q) v_2, \ldots) = T(v_1, v_2, \ldots).
\end{equation}

As shown in the Appendix, a tensor element $T(v_1, v_2, \ldots)$ vanishes unless
\begin{equation} \label{eqn:irrepcond}
\bigtimes_k c_k = \mathbb{I}
\end{equation}
where $c_k$ is the label of the irreducible representation that $v_k$ belongs to. Colloquially, this can be understood as conservation of charge at the tensor. As a direct consequence, if a fixed basis of eigenvectors of the generators is chosen, the multidimensional array $T_{i_1 i_2 i_3 \ldots}$ takes a block-sparse form, therefore reducing the number of non-zero parameters.

If we partition the indices to form two groups $\mathcal{I}_1, \mathcal{I}_2$, we can equivalently express the tensor as a linear operator
\begin{equation} \label{eqn:oprep}
\tilde{T}: \bigotimes_{k \in \mathcal{I}_1} \Hs_k \rightarrow \bigotimes_{k \in \mathcal{I}_2} \Hs_k
\end{equation}
where $(\otimes_{k \in \mathcal{I}_2} v_k)^\dagger \tilde{T} (\otimes_{k \in \mathcal{I}_1} v_k ) = T(v_1, v_2, v_3, \ldots)$. We refer to $\mathcal{I}_1$ as in-going and $\mathcal{I}_2$ as out-going indices. In a pictorial representation, we will associate arrows with the indices. What are the symmetry properties of this operator? As shown in the appendix, it commutes with the group action. Schur's Lemma then implies that for $x \in \bigotimes_{k \in \mathcal{I}_1} \Hs_k$ in the irreducible representation labeled $c$, $\tilde{T} x$ is also in the representation $c$. This is true for all possible partitions of the indices.

\subsection{Tensor contraction} \label{sct:contract}
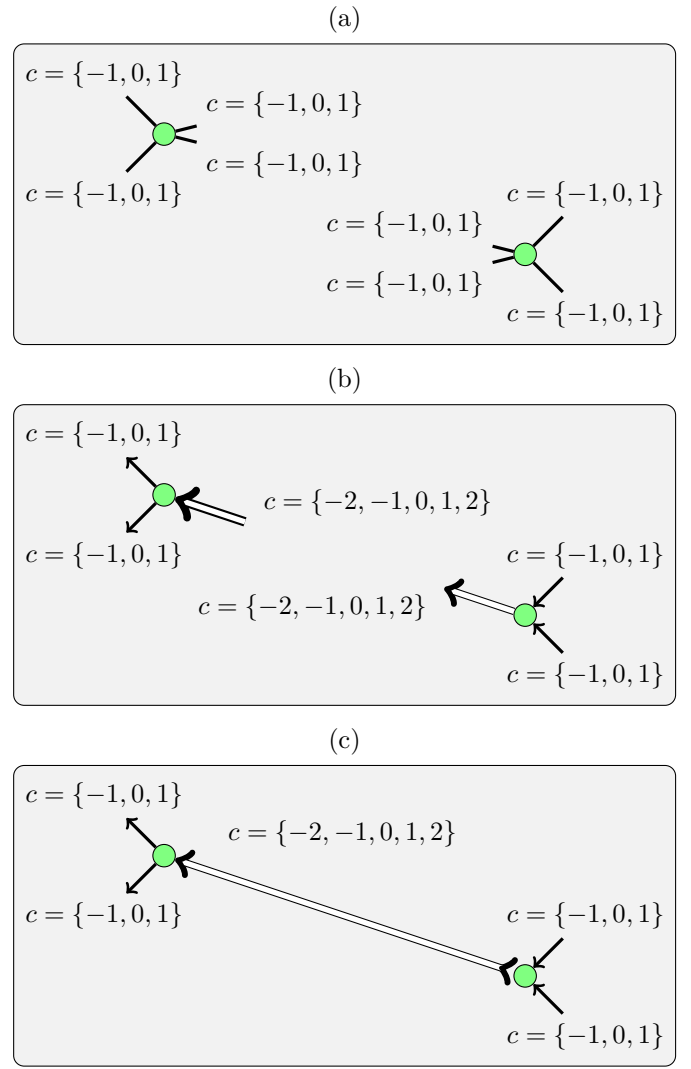
\begin{figure}[t]
	\beginpgfgraphicnamed{fig_tensor_ct}
	\begin{tikzpicture}[auto,scale=0.8]

	\node at (0,1.9) {(a)};

	\node[peps] (G0) at (-3,0) {};

	\node (leg01) at ($ (G0) + (-1,1) $) {$c = \lbrace -1,0,1 \rbrace$};
	\node (leg02) at ($ (G0) + (2,0.5) $) {$c = \lbrace -1,0,1 \rbrace$};
	\node (leg03) at ($ (G0) + (2,-0.5) $) {$c = \lbrace -1,0,1 \rbrace$};
	\node (leg04) at ($ (G0) + (-1,-1) $) {$c = \lbrace -1,0,1 \rbrace$};

	\draw[very thick] (G0) -- (leg01);
	\draw[very thick] (G0) -- (leg02);
	\draw[very thick] (G0) -- (leg03);
	\draw[very thick] (G0) -- (leg04);

	\node[peps] (G) at (3,-2) {};

	\node (leg1) at ($ (G) + (-2,0.5) $) {$c = \lbrace -1,0,1 \rbrace$};
	\node (leg2) at ($ (G) + (1,1) $) {$c = \lbrace -1,0,1 \rbrace$};
	\node (leg3) at ($ (G) + (1,-1) $) {$c = \lbrace -1,0,1 \rbrace$};
	\node (leg4) at ($ (G) + (-2,-0.5) $) {$c = \lbrace -1,0,1 \rbrace$};

	\draw[very thick] (G) -- (leg1);
	\draw[very thick] (G) -- (leg2);
	\draw[very thick] (G) -- (leg3);
	\draw[very thick] (G) -- (leg4);

	\begin{pgfonlayer}{background}
		\draw[rounded corners,fill=black!5] (-5.5,1.5) rectangle (5.5,-3.5);
	\end{pgfonlayer}



	\node at (0,1.9-6) {(b)};

	\node[peps] (G0) at (-3,-6) {};

	\node (leg01) at ($ (G0) + (-1,1) $) {$c = \lbrace -1,0,1 \rbrace$};
	\node (leg02) at ($ (G0) + (3/2,-1/2) $) {};
	\node (leg04) at ($ (G0) + (-1,-1) $) {$c = \lbrace -1,0,1 \rbrace$};

	\node[above right] at (leg02)  {$c = \lbrace -2,-1,0,1,2 \rbrace$};

	\draw[->,very thick] (G0) -- (leg01);
	\draw[<-,thick,double distance=2pt] (G0) -- (leg02);
	\draw[->,very thick] (G0) -- (leg04);

	\node[peps] (G) at (3,-8) {};

	\node (leg1) at ($ (G) + (-3/2,1/2) $) {};
	\node (leg2) at ($ (G) + (1,1) $) {$c = \lbrace -1,0,1 \rbrace$};
	\node (leg3) at ($ (G) + (1,-1) $) {$c = \lbrace -1,0,1 \rbrace$};

	\node[below left] at (leg1) {$c = \lbrace -2,-1,0,1,2 \rbrace $};

	\draw[->,double distance=2pt] (G) -- (leg1);
	\draw[<-,very thick] (G) -- (leg2);
	\draw[<-,very thick] (G) -- (leg3);

	\begin{pgfonlayer}{background}
		\draw[rounded corners,fill=black!5] (-5.5,1.5-6) rectangle (5.5,-3.5-6);
	\end{pgfonlayer}



	\node at (0,1.9-12) {(c)};

	\node[peps] (G0) at (-3,-12) {};

	\node (leg01) at ($ (G0) + (-1,1) $) {$c = \lbrace -1,0,1 \rbrace$};
	\node (leg02) at ($ (G0) + (5,0) $) {};
	\node (leg04) at ($ (G0) + (-1,-1) $) {$c = \lbrace -1,0,1 \rbrace$};

	\node[above left] at (leg02)  {$c = \lbrace -2,-1,0,1,2 \rbrace$};

	\draw[->,very thick] (G0) -- (leg01);
	\draw[->,very thick] (G0) -- (leg04);

	\node[peps] (G) at (3,-14) {};

	\node (leg1) at ($ (G) + (-5,0) $) {};
	\node (leg2) at ($ (G) + (1,1) $) {$c = \lbrace -1,0,1 \rbrace$};
	\node (leg3) at ($ (G) + (1,-1) $) {$c = \lbrace -1,0,1 \rbrace$};


	\draw[>->,double distance=2pt] (G) -- (G0);
	\draw[<-,very thick] (G) -- (leg2);
	\draw[<-,very thick] (G) -- (leg3);

	\begin{pgfonlayer}{background}
		\draw[rounded corners,fill=black!5] (-5.5,1.5-12) rectangle (5.5,-3.5-12);
	\end{pgfonlayer}

	\end{tikzpicture}
	\endpgfgraphicnamed

\caption{Pictorial representation of the contraction of two rank-4 tensors with $\uo$ symmetry. The steps are explained in detail in Section~\ref{sct:contract}. \label{fig:contraction} }
\end{figure}

The steps involved in the contraction of two rank-4 tensors over two indices are shown in Fig.~\ref{fig:contraction}. It is important to note at this point that in order to define these operations, only the charge calculus introduced in Section \ref{sct:nomenclature} is necessary. It is not necessary to know matrix representations of the group in all Hilbert spaces. This will allow us to introduce tensor networks with symmetries not just on physical, but also auxiliary bonds.

The steps of the contraction of two symmetric tensors are:
\begin{enumerate}[i)]
\item We first transform the tensors to operators of the form \eqref{eqn:oprep} (Fig.~\ref{fig:contraction} (a)--\ref{fig:contraction} (b)). The choices of in-going and out-going indices are dictated by the indices that are being contracted: on one tensor, those indices must be the in-going and on the other the out-going indices. The resulting operators, which are written as a matrix, have a block-diagonal structure.
\item The contraction is now equivalent to a matrix multiplication. The blocks must be contracted in such a way that the resulting tensor still satisfies \eqref{eqn:inv}. Therefore, we must match blocks in such a way that
\begin{equation}
c_{\text{in}} \times c_{\text{out}} = \mathbb{I}.
\end{equation}
\item The resulting tensor, Fig.~\ref{fig:contraction} (c), can be converted back to the form of Eqn. \eqref{eqn:tensor}.
\end{enumerate}

The conversion between the forms \eqref{eqn:tensor} and \eqref{eqn:oprep} also allows the definition of other linear algebra operations, such as singular value decomposition, eigenvalue decomposition based on the mapping to a matrix. All these share the block-diagonal structure.

\subsection{Symmetric PEPS}

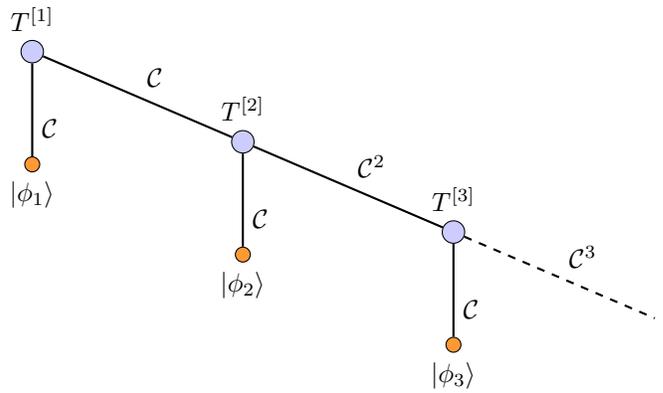
\begin{figure}[t]
	\beginpgfgraphicnamed{fig_symm_mps}
	\tikzset{gamma/.style={circle=2pt,draw=black!100,fill=blue!20,inner sep=3pt}}
	\tikzset{spin/.style={circle=1pt,draw=black!100,fill=orange!80,inner sep=2pt}}

	\begin{tikzpicture}[auto]

	\foreach \i / \j in {1/0,2/1,3/2}
	{
		\node (G\i) at (2.8*\j,-1.2*\j) [gamma] {};
		\node (S\i) at (2.8*\j,-1.5-1.2*\j) [spin] {};

		\node[above] (Gl\i) at (G\i.north) {$T^{[\i]}$};
		\node[below] at (S\i.south) {$| \phi_{\i} \rangle$};
		\draw[thick] (G\i) -- node[below right] {$\mathcal{C}$} (S\i);

	}

	\draw[thick] (G1) -- node[above right] {$\mathcal{C}$} (G2);
	\draw[thick] (G2) -- node[above right] {$\mathcal{C}^2$} (G3);

	\node (out) at (2.8*3,-3*1.2) {};
	\draw[thick,dashed] (G3) -- node [above right] {$\mathcal{C}^3$}(out);

	\end{tikzpicture}
	\endpgfgraphicnamed
\caption{End of a matrix product state invariant under some symmetry group $\mathcal{G}$. $|\phi_i\rangle$ denote physical states in the local Hilbert space $\Hs_{\text{loc}}$. By $\mathcal{C}$, we denote the set of charges associated with sectors in $\Hs_{\text{loc}}$, and by $\mathcal{C}^n$ the set of charges associated with sectors in $\bigotimes_{i=1}^n \Hs_{\text{loc}}$. The first auxiliary bond to the left simply carries the physical charges of the first site. For the second bond, all possible fusion outcomes of charges on the first auxiliary bond with the physical charges have to be considered. This can be continued up to the middle of the chain, such that each auxiliary bond carries the possible combinations of charges to the left. Joining such a state with its reflection will yield a finite symmetric MPS. \label{fig:symmmps} }
\end{figure}

As a simple example of a tensor network, the construction of a matrix-product state invariant under some symmetry group $\mathcal{G}$ is shown in Fig. \ref{fig:symmmps}. On each bond $i$ of the tensor network, we have a set of charges $\mathcal{C}_i$. For the bonds connecting to physical degrees of freedom, this set of charges is fixed by the physical Hilbert space. In the case of a finite MPS with open boundary conditions, the set of charges possible on an auxiliary bond is unique and has a well-defined physical meaning: if one were to consider, e.g., a system with particle number conservation, the allowed symmetry sectors on each auxiliary bond in the construction in Fig. \ref{fig:symmmps} are the possible particle numbers to the left part of the chain. In general, the set of allowed charges corresponds to the possible fusion outcomes of all physical charges to the left. In a finite system, a quantum number sector can be selected by appropriately fixing the allowed charges at the right end of the chain.

In the case of a PEPS, a unique identification of the charges on an auxiliary bond with the fusion outcomes of a specific region cannot be made. It is therefore not possible to determine uniquely which symmetry sectors must be kept on the auxiliary bonds. While for finite groups, it is usually computationally possible to allow all charge sectors, some choice has to be made in the case of infinite groups. It will therefore be one of the main purposes of this paper to verify that i) for finite and infinite groups, one obtains a good approximation to the ground state by using a PEPS constructed from symmetric tensors, ii) for infinite groups, a reasonable approximation is obtained for computationally feasible choices of the symmetry sectors.

\begin{figure}[t]
	\beginpgfgraphicnamed{fig_symm_peps}

	\tikzset{corner/.style={rectangle=2pt,draw=black!100,fill=red!30,inner sep=3pt}}
	\tikzset{gamma/.style={circle=2pt,draw=black!100,fill=blue!20,inner sep=3pt}}
	\tikzset{spin/.style={circle=1pt,draw=black!100,fill=orange!80,inner sep=2pt}}
	\tikzset{Mbond/.style={thick,blue,dashed}}

	\begin{tikzpicture}[auto]

	\foreach \i / \j in {0/0,1/0,2/0,0/-1}
	{
		\node(C\i\j) at (2*\i,2*\j) [corner] {};
	}

	\foreach \i / \j in {1/-1,2/-1}
	{
		\node(G\i\j) at (2*\i-0.4,2*\j+0.4) [gamma] {};
		\node(Gp\i\j) at (2*\i+0.4,2*\j-0.4) [gamma] {};
		\node(S\i\j) at (2*\i,2*\j) [spin] {};
	        \node[above left,blue] at (G\i\j) {$T_\i^*$};
	        \node[below right,blue] at (Gp\i\j) {$T_\i$};
	}

	\draw[thick] (G1-1) -- (S1-1);
	\draw[thick] (S1-1) -- (Gp1-1);
	\draw[thick] (G2-1) -- (S2-1);
	\draw[thick] (S2-1) -- (Gp2-1);

	\draw[Mbond] (G1-1) -- (G2-1);
	\draw[Mbond] (Gp1-1) -- (Gp2-1);

	\draw[thick,red] (C00) -- node[above] {$\mathcal{C}_{\text{corner}}$} (C10);
	\draw[thick,red] (C10) -- (C20);
	\draw[thick,red] (C00) -- (C0-1);

	\draw[Mbond] (C10) -- (G1-1);
	\draw[Mbond] (C10) -- node[right] {$\mathcal{C}_{\text{aux}}$} (Gp1-1);
	\draw[Mbond] (C20) -- (G2-1);
	\draw[Mbond] (C20) -- (Gp2-1);
	\draw[Mbond] (C0-1) -- (G1-1);
	\draw[Mbond] (C0-1) -- (Gp1-1);

	\draw[thick,red] (C0-1) -- ($ (C0-1) + (0,-2) $);
	\draw[thick,red] (C20) -- ($ (C20) + (2,0) $);

	\foreach \i in {G1-1,Gp1-1,G2-1,Gp2-1}
	{
		\draw[Mbond] (\i) -- ($ (\i) + (0,-2) $);
	}

	\foreach \i in {G2-1,Gp2-1}
	{
		\draw[Mbond] (\i) -- ($ (\i) + (2,0) $);
	}

	\end{tikzpicture}
	\endpgfgraphicnamed

\caption{Corner of a symmetric PEPS state with an environment as in the directional corner transfer matrix method.\cite{orus2009-1} Here, the blue circles denote tensors $T_i$ of the ansatz state and their conjugate $T_i^*$, the red squares denote tensors of the corner transfer matrix and the orange circles represent single-site operators acting on the physical index of the PEPS tensor. In the infinite case, in general there are three sets of charges involved: i) the physical charges $\mathcal{C}_{\text{phys}}$ carried on the black, solid lines in the figure, ii) the auxiliary charges $\mathcal{C}_{\text{aux}}$ on the blue, dashed bonds, and iii) the charges carried on the bonds of the environment $\mathcal{C}_{\text{corner}}$. This reflects the three independent bond dimensions involved in a PEPS: the physical dimension $d$, the bond dimension $M$ and the environment dimension $\chi$. Usually, $M > d$ and $\chi \sim M^2$. Therefore, $\mathcal{C}_{\text{phys}} \subset \mathcal{C}_{\text{aux}} \subset \mathcal{C}_{\text{corner}}$. In principle, all charges could depend on the location in the PEPS or the environment. \label{fig:symmpeps} }
\end{figure}
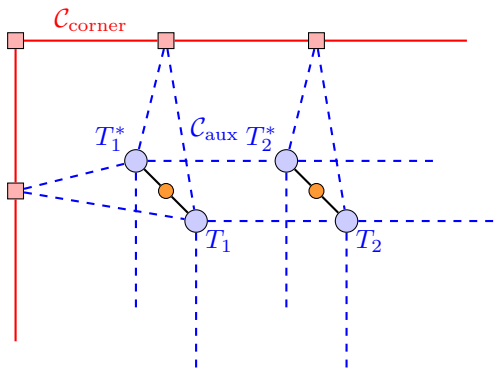

To understand the nature of the approximation introduced by truncating the set of allowed quantum numbers, consider the expansion of a state $|\Psi\rangle = \sum_{|\phi\rangle} c(\phi) |\phi\rangle$. Using a tensor network ansatz amounts to representing all coefficients $c(\phi)$ by a trace over a tensor network, which will represent the low-entanglement subspace of the full Hilbert space efficiently. In principle, all basis states are allowed and could have non-vanishing weight. Imposing restrictions on the quantum numbers, on the other hand, amounts to a restriction on the allowed basis states: the sum does not run over the full basis $\lbrace |\phi\rangle\rbrace$, but only a subset of states compatible with the allowed quantum numbers.

In addition to the charge sectors on each bond, the number of states in each sector has to be chosen. In principle, this could differ between all sectors on one bond and between bonds. The situation becomes more involved since even for a translational invariant PEPS, several different sets of charges have to be considered, as shown in Fig. \ref{fig:symmpeps}. For the purpose of this paper, we make the simplification that we choose the charges to be the same on all equivalent bonds of the lattice and the environment states. Additionally, for the case of finite groups, we choose the number of states in each sector the same for all equivalent bonds.

Two points require special attention when applying symmetric PEPS to infinite lattices: i) On infinite lattices, the ansatz is restricted to states that globally fall into the sector of the identity charge. For example, using the $U(1)$ symmetry of a spin-$\frac{1}{2}$ system described in Section~\ref{sct:u1}, only states with vanishing total magnetization can be studied. In the case of particle number conservation, an appropriate choice of charges would have to be taken to enforce the desired filling fraction. On finite lattices, however, selecting specific quantum number sectors is possible also in the PEPS construction by adding an external bond carrying the total charge of the system to one of the tensors that make up the PEPS. ii) Since our construction assumes that the state has a well-defined global quantum number, systems that spontaneously break the symmetry that is being exploited cannot be represented. If the possibility of a spontaneous symmetry breaking is present in the system being studied, the results should therefore be checked against calculations without enforcing the symmetry.

\subsection{Implementation}

In this section, we will outline a few details of our implementation. The most important operations on tensors include i) contraction, ii) singular-value decomposition, and iii) eigenvalue decomposition of tensors. In particular for the last two operations, very efficient implementations exist for matrices and it is advisable to make use of these. The contraction could in principle be implemented directly as a summation, however it turns out to be favorable to map it to matrix multiplication and make use of existing, optimized implementations.

The most important operation therefore is the mapping between a symmetric tensor and a block-sparse matrix, i.e. between the form \eqref{eqn:tensor} and \eqref{eqn:oprep}, with several tensor indices grouped to form the left and right indices of the matrix. Since this operation, like a matrix transpose, scales like $\mathcal{O}(N)$, with $N$ the number of elements in the matrix, it is subleading compared to contraction, singular-value decomposition, etc., which all scale roughly as $\mathcal{O}(N^{3/2})$.

For the conversion between tensors and matrices, one could either calculate the correspondence between the location of an element in the tensor and in the matrix on the fly or compute it once and store it in memory along with the tensor (precomputation). This is explained in some detail in Ref. \onlinecite{singh2010}. We choose not to use precomputation for several reasons: a) the overhead in memory usage may be significant, b) memory bandwidth is one of the bottlenecks of tensor-network state simulations and should therefore be minimized, c) the structure of tensors, in particular for groups such as $\uo$ where the number of sectors is chosen dynamically, may vary between each iteration of the algorithm, d) the overhead of calculating the tensor structure on the fly is negligible if implemented efficiently in a compiled language such as \texttt{C++}.

\section{Results}

\begin{figure}
\includegraphics[width=\columnwidth]{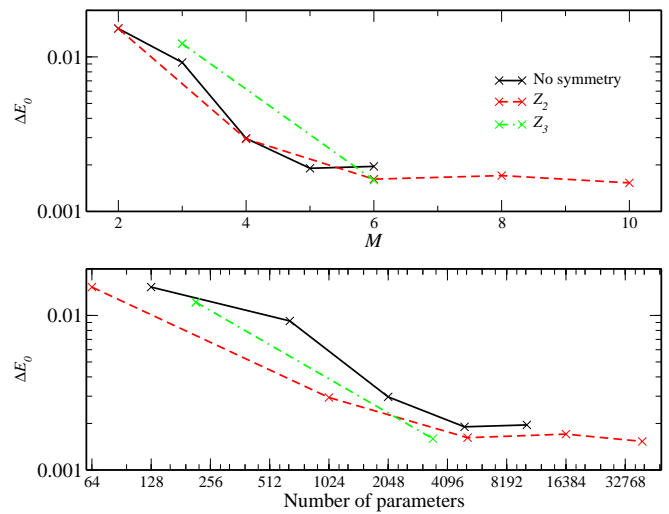}
\caption{The first panel shows how the relative error in the energy decreases as $M$ is increased, for different choices of the symmetry group. The second panel shows the same data versus the number of variational parameters in the state (note the logarithmic scale on both axes). Clearly, the fact that the relative errors are similar between symmetry groups for a given $M$ shows that reducing the number of parameters and the computation time by using larger (finite) symmetry groups does not lead to any loss in accuracy. \label{fig:HB_Zq} }
\end{figure}

\begin{figure}
\includegraphics[width=\columnwidth]{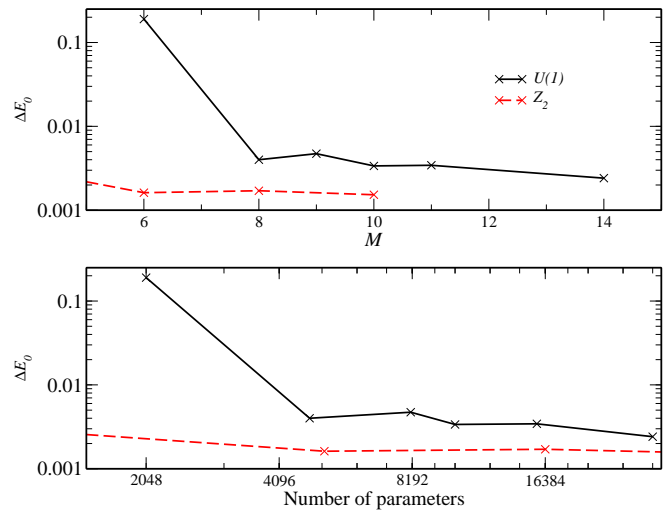}
\caption{Relative error in the ground state energy of the 2d Heisenberg model with a $\uo$-symmetric PEPS, as a function of i) the total bond dimension, ii) the number of parameters. The choice of sectors and dimensions is shown in Table~\ref{tab:uo}. \label{fig:HB_U1} We show results with $\zz$ symmetry for comparison.}
\end{figure}

It has been demonstrated in Ref. \onlinecite{bauer2009} that the spin-$\frac{1}{2}$ Heisenberg model,
\begin{equation}
H = \sum_{\langle i,j \rangle} \vec{S}_i \cdot \vec{S}_j,
\end{equation}
where $\vec{S}_i$ is the spin-$\frac{1}{2}$ operator at site $i$ and the summation runs over pairs of nearest neighbors, is a difficult test case for tensor network methods. As in Ref.~\onlinecite{bauer2009}, we will work on an infinite square lattice. This is due to strong fluctuations around the N\'{e}el state, which reduce local magnetic moments significantly. We will use the Heisenberg model as a benchmark case here and compare to precision Monte Carlo calculations.\cite{sandvik1997,sandvik1999}

All results in this section were obtained using the simplified update described in Refs.~\onlinecite{jiang2008,corboz2010}. In this update scheme, long-range correlations are effectively taken into account by introducing weights on the auxiliary bonds of the PEPS. Imaginary time evolution is then performed locally, determining new tensors and weights bond after bond. While no formal justification can be given for the weights, the accuracy of the algorithm applied to systems away from criticality turns out to be only slightly less than an update scheme that takes correlations into account more rigorously. The advantage of the simplified algorithm lies in the much better performance and robustness against numerical instabilities.

For the imaginary time evolution, a Suzuki-Trotter decomposition of the evolution operator has to be performed. Since numerical errors do not accumulate in imaginary time evolution, we can reduce the time discretization during our simulation and completely suppress discretization errors. To extract expectation values, we use the directional corner transfer matrix approach of Ref.~\onlinecite{orus2009-1}. We use an ansatz with 4 independent tensors in a $2 \times 2$ unit cell.

\subsection{Finite groups}

The results we obtain for the Heisenberg model with finite symmetry groups $\zz$ and $\mathbb{Z}_3$ are shown in Fig. \ref{fig:HB_Zq} as a function of the total bond dimension on the bonds of the state and as a function of the total number of variational parameters of the state (note the logarithmic scale in this case). For comparison, we show results obtained with a non-symmetric PEPS, but with the same simplified update scheme. We choose the number of states equal in each sector, hence $M = q\cdot n$. We also make the same choice on all bonds of the PEPS. We keep up to 36 states in the renormalization of the corner transfer matrix.

For $n > 1$, that is with a non-trivial dimension in each symmetry sector on the auxiliary bonds, the energies obtained with the symmetric PEPS are comparable to those obtained without symmetry for the same bond dimension. This demonstrates that the approximation introduced by restricting the structure of the tensors is valid and does not affect the accuracy. Since all matrix operations decompose into $q$ blocks, we can expect a speedup of $\ord{q^3}$ of the algorithm. In terms of the number of variational parameters, a significant improvement is achieved: with $\zz$ symmetry, only half the number of variational parameters is necessary. With $\zq$ symmetry, the reduction is even stronger. This may be advantageous particularly if a direct energy minimization algorithm is applied instead of the imaginary time evolution.

In some cases, the energy of the symmetric state falls below the energy of the non-symmetric state. This must be attributed to trapping in local minima, which seems more likely in the case of a non-symmetric PEPS with more variational parameters.

\subsection{$\uo$}

\begin{table}
\begin{tabular}{|c|c|c|c|c|}
\hline $n$ &$M$ &$M_c$ &Parameters& Comparison $\zz$ \\ \hline
3 &6 &2-2-2 &2048 &5184 \\
5 &8 &1-2-2-2-1 &4800 &16384 \\
5 &9 &1-2-3-2-1 &8128 &26244 \\
5 &10 &2-2-2-2-2 &10240 &40000 \\
5 &11 &2-2-3-2-2 &15680 &58564 \\
7 &14 &2-2-2-2-2-2-2 &28672 &153664 \\ \hline
\end{tabular}
\caption{The table shows the choices of symmetry-sector dimensions for the $\uo$ symmetry in the Heisenberg model. The columns contain i) the number of sectors associated with quantum numbers $S_z = -(n-1)/2 \ldots (n-1)/2$, ii) the total bond dimension, iii) the size of each sector, iv) the total number of parameters of the state, v) the total number of parameters in a $\zz$-symmetric state with the same total bond dimension.  \label{tab:uo} }
\end{table}

While for the finite groups considered so far, we could simply keep all allowed sectors of the symmetry on each auxiliary bond, some choice must be made for the infinite group $\uo$. Furthermore, we have to choose the dimension within each symmetry sector -- due to the large number of sectors, it is generally not efficient to keep it the same in all sectors, as we did for finite groups. However, given the fast growth of computational cost with the bond dimension, only a few choices are possible. The choices we considered are listed in the table in Fig.~\ref{tab:uo}. It should be noted that for equal total bond dimension $M$, a state with more symmetry sectors of smaller dimension is computationally less expensive since all matrix computations can be split into more blocks. This allows us to study states with very large bond dimension up to $M=14$, which would be intractable otherwise.

Results obtained with the above choices are shown in Fig.~\ref{fig:HB_U1}. The accuracy for a given bond dimension is worse than with the finite group $\zz$; even for the very large bond dimensions studied with $\uo$ symmetry, the accuracy does not reach the level of the finite symmetry groups. This is a clear signature that the approximation we made by imposing a $\uo$-symmetric structure on the tensors and picking only a few allowed sectors of the symmetry limits the accuracy of the simulations. One has to keep in mind, however, that the number of variational parameters is reduced much more strongly than in the case of finite groups, as shown in the last column of Table~\ref{tab:uo}.

\section{Conclusion}

We have explained a formalism for introducing Abelian symmetries into tensor network state algorithms. The formalism relies only on fusion properties of irreducible representations and is therefore easily applied to a large class of symmetry groups. The formalism can be applied to any tensor network state algorithm; for this paper, we have restricted ourselves to simulations with projected entangled-pair states.

Since the implementation requires additional approximations, benchmark calculations confirming the validity of the approach are required. This is particularly important in the case of $\uo$, where restrictions on the allowed quantum numbers have to be introduced. In order to assess the validity, we have applied our method to the spin-$\frac{1}{2}$ Heisenberg model on an infinite square lattice with the symmetry groups $\zq$ for $q=2,3$ and $\uo$.

Our results for the finite groups show that no accuracy is lost due to the symmetric decomposition of tensors. At the same time, the number of variational parameters and the computational effort is significantly reduced. We therefore expect that exploiting these symmetries will become very useful in the context of tensor networks states. As a future direction of research, a decomposition where the number of states in each symmetry sector is not equal could be considered, which may lead to even better accuracy for a given bond dimension.

In the case of the continuous symmetry group $\uo$, we were able to achieve much larger bond dimensions. Nevertheless, the accuracy does not reach the level that can be obtained with finite symmetry groups. We expect, however, that if a sufficiently large number of symmetry sectors is taken into account, the accuracy will eventually become comparable to the non-symmetric case. Further research is required to understand this, in particular how this behaves for different models such as bosonic models with particle number conservation. Also, a scheme that automatically picks the relevant symmetry sectors on the auxiliary bonds and in the environment tensors without strong dependence on the initial state may be very useful.

We acknowledge useful discussions with G. Vidal. Simulations were performed on the Brutus cluster at ETH Zurich.

\section{Appendix} \label{sct:appdx}

In this Appendix, we show in detail some calculations relevant to the discussion in Sect.~\ref{sct:symmtensor}, in particular symmetry properties of a tensor of the form~\eqref{eqn:tensor}.

To simplify the notation, we consider the case of a tensor of rank 2, where the Hilbert spaces are taken to be equal:
\begin{eqnarray}
T &:& \Hs \otimes \Hs \rightarrow \mathbb{C} \\
\tilde{T} &:& \Hs \rightarrow \Hs,
\end{eqnarray}
where for $x,y \in \Hs$ we have $T(x,y) = y^\dagger \tilde{T} x$. Also, let $U$ be a unitary representation of the group in $\Hs$.

We would like to show the following equivalence:
\begin{eqnarray}
(a)\ \ \  &\ &[\tilde{T},U] = 0\\
(b)\ \ \  &\Leftrightarrow& T(U x, U y) = T(x,y). \nonumber
\end{eqnarray}
First, we show that $(b)$ follows from $(a)$:
\begin{eqnarray}
T(U x, U y) &=& y^\dagger U^\dagger \tilde{T} U x \\
\ &=& y^\dagger U^\dagger U \tilde{T} x \nonumber \\
\ &=& T(x,y)  \nonumber
\end{eqnarray}
Secondly, we show that $(a)$ follows from $(b)$:
\begin{eqnarray}
y^\dagger [T,U] x &=& y^\dagger T U x - y^\dagger U T x \\
\ &=& T(U x, y) - T(x, U^\dagger y) \nonumber \\
\ &=& T(U^\dagger U x, U^\dagger y) - T(x, U^\dagger y) \nonumber \\
\ &=& 0 \nonumber
\end{eqnarray}

The above can easily be generalized for all operators of the form \eqref{eqn:oprep}.

We now want to show the validity of \eqref{eqn:irrepcond}. Let $q \in \mathcal{G}$. Then,
\begin{eqnarray}
\ &\ &T(U(q) v_1, U(q) v_2, \ldots) \\
\  &=& T(\nu(q, c_1) v_1, \nu(q, c_2) v_2, \ldots) \nonumber \\
\ &=& \nu(q, c_1) \nu(q, c_2) \ldots T(v_1, v_2, \ldots) \nonumber \\
\ &=& \nu(q, \bigtimes c_k) T(v_1, v_2, \ldots) \nonumber \\
\ &=& T(v_1, v_2, \ldots) \nonumber
\end{eqnarray}
where we have used Eqns.~\eqref{eqn:inv} and \eqref{eqn:numult}. Therefore, $\nu(q, \bigtimes c_k) = 1$ and $\bigtimes c_k = \mathbb{I}$.

\bibliography{bibliography}{}

\begin{thebibliography}{10}%
\makeatletter
\providecommand \@ifxundefined [1]{%
 \ifx #1\undefined \expandafter \@firstoftwo
 \else \expandafter \@secondoftwo
\fi
}%
\providecommand \@ifnum [1]{%
 \ifnum #1\expandafter \@firstoftwo
 \else \expandafter \@secondoftwo
\fi
}%
\providecommand \enquote [1]{``#1''}%
\providecommand \bibnamefont  [1]{#1}%
\providecommand \bibfnamefont [1]{#1}%
\providecommand \citenamefont [1]{#1}%
\providecommand\href[0]{\@sanitize\@href}%
\providecommand\@href[1]{\endgroup\@@startlink{#1}\endgroup\@@href}%
\providecommand\@@href[1]{#1\@@endlink}%
\providecommand \@sanitize [0]{\begingroup\catcode`\&12\catcode`\#12\relax}%
\@ifxundefined \pdfoutput {\@firstoftwo}{%
 \@ifnum{\z@=\pdfoutput}{\@firstoftwo}{\@secondoftwo}%
}{%
 \providecommand\@@startlink[1]{\leavevmode\special{html:<a href="#1">}}%
 \providecommand\@@endlink[0]{\special{html:</a>}}%
}{%
 \providecommand\@@startlink[1]{%
  \leavevmode
  \pdfstartlink
   attr{/Border[0 0 1 ]/H/I/C[0 1 1]}%
   user{/Subtype/Link/A<</Type/Action/S/URI/URI(#1)>>}%
  \relax
 }%
 \providecommand\@@endlink[0]{\pdfendlink}%
}%
\providecommand \url  [0]{\begingroup\@sanitize \@url }%
\providecommand \@url [1]{\endgroup\@href {#1}{\urlprefix}}%
\providecommand \urlprefix [0]{URL }%
\providecommand \Eprint[0]{\href }%
\@ifxundefined \urlstyle {%
  \providecommand \doi [1]{doi:\discretionary{}{}{}#1}%
}{%
  \providecommand \doi [0]{doi:\discretionary{}{}{}\begingroup
  \urlstyle{rm}\Url }%
}%
\providecommand \doibase [0]{http://dx.doi.org/}%
\providecommand \Doi[1]{\href{\doibase#1}}%
\providecommand \bibAnnote [3]{%
  \BibitemShut{#1}%
  \begin{quotation}\noindent
    \textsc{Key:}\ #2\\\textsc{Annotation:}\ #3%
  \end{quotation}%
}%
\providecommand \bibAnnoteFile [2]{%
  \IfFileExists{#2}{\bibAnnote {#1} {#2} {\input{#2}}}{}%
}%
\providecommand \typeout [0]{\immediate \write \m@ne }%
\providecommand \selectlanguage [0]{\@gobble}%
\providecommand \bibinfo [0]{\@secondoftwo}%
\providecommand \bibfield [0]{\@secondoftwo}%
\providecommand \translation [1]{[#1]}%
\providecommand \BibitemOpen[0]{}%
\providecommand \bibitemStop [0]{}%
\providecommand \bibitemNoStop [0]{.\EOS\space}%
\providecommand \EOS [0]{\spacefactor3000\relax}%
\providecommand \BibitemShut [1]{\csname bibitem#1\endcsname}%
\bibitem{white1992}%
  \BibitemOpen
  \bibfield{author}{%
  \bibinfo {author} {\bibfnamefont{S.~R.}\ \bibnamefont{White}},\ }%
  \bibfield{journal}{%
  \bibinfo {journal} {Phys. Rev. Lett.}\ }%
  \textbf{\bibinfo {volume} {69}},\ \bibinfo {pages} {2863} (\bibinfo {year}
  {1992})%
  \bibAnnoteFile{NoStop}{white1992}%
\bibitem{ostlund1995}%
  \BibitemOpen
  \bibfield{author}{%
  \bibinfo {author} {\bibfnamefont{S.}~\bibnamefont{{\"O}stlund}}\ and\
  \bibinfo {author} {\bibfnamefont{S.}~\bibnamefont{Rommer}},\ }%
  \bibfield{journal}{%
  \bibinfo {journal} {Phys. Rev. Lett.}\ }%
  \textbf{\bibinfo {volume} {75}},\ \bibinfo {pages} {3537} (\bibinfo {year}
  {1995})%
  \bibAnnoteFile{NoStop}{ostlund1995}%
\bibitem{verstraete2004}%
  \BibitemOpen
  \bibfield{author}{%
  \bibinfo {author} {\bibfnamefont{F.}~\bibnamefont{Verstraete}}\ and\ \bibinfo
  {author} {\bibfnamefont{J.~I.}\ \bibnamefont{Cirac}},\ }%
  \bibfield{journal}{%
  \bibinfo {journal} {Preprint}}%
   (\bibinfo {year} {2004}),\ \bibinfo {note} {arXiv:0407.066}%
  \bibAnnoteFile{NoStop}{verstraete2004}%
\bibitem{sierra1998}%
  \BibitemOpen
  \bibfield{author}{%
  \bibinfo {author} {\bibfnamefont{G.}~\bibnamefont{Sierra}}\ and\ \bibinfo
  {author} {\bibfnamefont{M.~A.}\ \bibnamefont{Martin-Delgado}},\ }%
  \bibfield{journal}{%
  \bibinfo {journal} {Preprint}}%
   (\bibinfo {year} {1998}),\ \bibinfo {note} {arXiv:cond-mat/9811170v3}%
  \bibAnnoteFile{NoStop}{sierra1998}%
\bibitem{nishino1998}%
  \BibitemOpen
  \bibfield{author}{%
  \bibinfo {author} {\bibfnamefont{T.}~\bibnamefont{Nishino}}\ and\ \bibinfo
  {author} {\bibfnamefont{K.}~\bibnamefont{Okunishi}},\ }%
  \bibfield{journal}{%
  \bibinfo {journal} {Journal of the Physical Society of Japan}\ }%
  \textbf{\bibinfo {volume} {67}},\ \bibinfo {pages} {3066} (\bibinfo {year}
  {1998})%
  \bibAnnoteFile{NoStop}{nishino1998}%
\bibitem{nishino2000}%
  \BibitemOpen
  \bibfield{author}{%
  \bibinfo {author} {\bibfnamefont{T.}~\bibnamefont{Nishino}}, \bibinfo
  {author} {\bibfnamefont{K.}~\bibnamefont{Okunushi}}, \bibinfo {author}
  {\bibfnamefont{Y.}~\bibnamefont{Hieida}}, \bibinfo {author}
  {\bibfnamefont{N.}~\bibnamefont{Maeshima}},\ and\ \bibinfo {author}
  {\bibfnamefont{Y.}~\bibnamefont{Akutsu}},\ }%
  \bibfield{journal}{%
  \bibinfo {journal} {Nucl. Phys. B}\ }%
  \textbf{\bibinfo {volume} {575}},\ \bibinfo {pages} {504} (\bibinfo {year}
  {2000})%
  \bibAnnoteFile{NoStop}{nishino2000}%
\bibitem{nishio2004}%
  \BibitemOpen
  \bibfield{author}{%
  \bibinfo {author} {\bibfnamefont{Y.}~\bibnamefont{Nishio}}, \bibinfo {author}
  {\bibfnamefont{N.}~\bibnamefont{Maeshima}}, \bibinfo {author}
  {\bibfnamefont{A.}~\bibnamefont{Gendiar}},\ and\ \bibinfo {author}
  {\bibfnamefont{T.}~\bibnamefont{Nishino}},\ }%
  \bibfield{journal}{%
  \bibinfo {journal} {Preprint}}%
   (\bibinfo {year} {2004}),\ \bibinfo {note} {arXiv:cond-mat/0401115v1}%
  \bibAnnoteFile{NoStop}{nishio2004}%
\bibitem{murg2007}%
  \BibitemOpen
  \bibfield{author}{%
  \bibinfo {author} {\bibfnamefont{V.}~\bibnamefont{Murg}}, \bibinfo {author}
  {\bibfnamefont{F.}~\bibnamefont{Verstraete}},\ and\ \bibinfo {author}
  {\bibfnamefont{J.~I.}\ \bibnamefont{Cirac}},\ }%
  \bibfield{journal}{%
  \bibinfo {journal} {Phys. Rev. A}\ }%
  \textbf{\bibinfo {volume} {75}},\ \bibinfo {pages} {033605} (\bibinfo {year}
  {2007})%
  \bibAnnoteFile{NoStop}{murg2007}%
\bibitem{pchen2009}%
  \BibitemOpen
  \bibfield{author}{%
  \bibinfo {author} {\bibfnamefont{P.}~\bibnamefont{Chen}}, \bibinfo {author}
  {\bibfnamefont{C.-Y.}\ \bibnamefont{Lai}},\ and\ \bibinfo {author}
  {\bibfnamefont{M.-F.}\ \bibnamefont{Yang}},\ }%
  \bibfield{journal}{%
  \bibinfo {journal} {J. Stat. Mech.}\ }%
  \textbf{\bibinfo {volume} {2009}},\ \bibinfo {pages} {P10001} (\bibinfo
  {year} {2009})%
  \bibAnnoteFile{NoStop}{pchen2009}%
\bibitem{nishino2001}%
  \BibitemOpen
  \bibfield{author}{%
  \bibinfo {author} {\bibfnamefont{T.}~\bibnamefont{Nishino}}, \bibinfo
  {author} {\bibfnamefont{Y.}~\bibnamefont{Hieida}}, \bibinfo {author}
  {\bibfnamefont{K.}~\bibnamefont{Okunushi}}, \bibinfo {author}
  {\bibfnamefont{N.}~\bibnamefont{Maeshima}}, \bibinfo {author}
  {\bibfnamefont{Y.}~\bibnamefont{Akutsu}},\ and\ \bibinfo {author}
  {\bibfnamefont{A.}~\bibnamefont{Gendiar}},\ }%
  \bibfield{journal}{%
  \bibinfo {journal} {Progr. Theor. Phys.}\ }%
  \textbf{\bibinfo {volume} {105}},\ \bibinfo {pages} {409} (\bibinfo {year}
  {2001})%
  \bibAnnoteFile{NoStop}{nishino2001}%
\bibitem{maeshima2001}%
  \BibitemOpen
  \bibfield{author}{%
  \bibinfo {author} {\bibfnamefont{N.}~\bibnamefont{Maeshima}}, \bibinfo
  {author} {\bibfnamefont{Y.}~\bibnamefont{Hieida}}, \bibinfo {author}
  {\bibfnamefont{Y.}~\bibnamefont{Akutsu}}, \bibinfo {author}
  {\bibfnamefont{T.}~\bibnamefont{Nishino}},\ and\ \bibinfo {author}
  {\bibfnamefont{K.}~\bibnamefont{Okunishi}},\ }%
  \bibfield{journal}{%
  \bibinfo {journal} {Phys. Rev. E}\ }%
  \textbf{\bibinfo {volume} {64}},\ \bibinfo {pages} {016705} (\bibinfo {year}
  {2001})%
  \bibAnnoteFile{NoStop}{maeshima2001}%
\bibitem{gendiar2003}%
  \BibitemOpen
  \bibfield{author}{%
  \bibinfo {author} {\bibfnamefont{A.}~\bibnamefont{Gendiar}}, \bibinfo
  {author} {\bibfnamefont{N.}~\bibnamefont{Maeshima}},\ and\ \bibinfo {author}
  {\bibfnamefont{T.}~\bibnamefont{Nishino}},\ }%
  \bibfield{journal}{%
  \bibinfo {journal} {Progr. Theor. Phys.}\ }%
  \textbf{\bibinfo {volume} {110}},\ \bibinfo {pages} {691} (\bibinfo {year}
  {2003})%
  \bibAnnoteFile{NoStop}{gendiar2003}%
\bibitem{gendiar2005}%
  \BibitemOpen
  \bibfield{author}{%
  \bibinfo {author} {\bibfnamefont{A.}~\bibnamefont{Gendiar}}, \bibinfo
  {author} {\bibfnamefont{T.}~\bibnamefont{Nishino}},\ and\ \bibinfo {author}
  {\bibfnamefont{R.}~\bibnamefont{Derian}},\ }%
  \bibfield{journal}{%
  \bibinfo {journal} {Acta Phys. Slov.}\ }%
  \textbf{\bibinfo {volume} {55}},\ \bibinfo {pages} {141} (\bibinfo {year}
  {2005})%
  \bibAnnoteFile{NoStop}{gendiar2005}%
\bibitem{isacsson2006}%
  \BibitemOpen
  \bibfield{author}{%
  \bibinfo {author} {\bibfnamefont{A.}~\bibnamefont{Isacsson}}\ and\ \bibinfo
  {author} {\bibfnamefont{O.~F.}\ \bibnamefont{Syljuasen}},\ }%
  \bibfield{journal}{%
  \bibinfo {journal} {Phys. Rev. E}\ }%
  \textbf{\bibinfo {volume} {74}},\ \bibinfo {pages} {026701} (\bibinfo {year}
  {2006})%
  \bibAnnoteFile{NoStop}{isacsson2006}%
\bibitem{jiang2008}%
  \BibitemOpen
  \bibfield{author}{%
  \bibinfo {author} {\bibfnamefont{H.~C.}\ \bibnamefont{Jiang}}, \bibinfo
  {author} {\bibfnamefont{Z.~Y.}\ \bibnamefont{Weng}},\ and\ \bibinfo {author}
  {\bibfnamefont{T.}~\bibnamefont{Xiang}},\ }%
  \bibfield{journal}{%
  \bibinfo {journal} {Phys. Rev. Lett.}\ }%
  \textbf{\bibinfo {volume} {101}},\ \bibinfo {pages} {090603} (\bibinfo {year}
  {2008})%
  \bibAnnoteFile{NoStop}{jiang2008}%
\bibitem{gu2008}%
  \BibitemOpen
  \bibfield{author}{%
  \bibinfo {author} {\bibfnamefont{Z.-C.}\ \bibnamefont{Gu}}, \bibinfo {author}
  {\bibfnamefont{M.}~\bibnamefont{Levin}},\ and\ \bibinfo {author}
  {\bibfnamefont{X.-G.}\ \bibnamefont{Wen}},\ }%
  \bibfield{journal}{%
  \bibinfo {journal} {Phys. Rev. B}\ }%
  \textbf{\bibinfo {volume} {78}},\ \bibinfo {pages} {205116} (\bibinfo {year}
  {2008})%
  \bibAnnoteFile{NoStop}{gu2008}%
\bibitem{murg2009}%
  \BibitemOpen
  \bibfield{author}{%
  \bibinfo {author} {\bibfnamefont{V.}~\bibnamefont{Murg}}, \bibinfo {author}
  {\bibfnamefont{F.}~\bibnamefont{Verstraete}},\ and\ \bibinfo {author}
  {\bibfnamefont{J.~I.}\ \bibnamefont{Cirac}},\ }%
  \bibfield{journal}{%
  \bibinfo {journal} {Preprint}}%
   (\bibinfo {year} {2009}),\ \bibinfo {note} {arXiv:0901.2019v1}%
  \bibAnnoteFile{NoStop}{murg2009}%
\bibitem{xie2009}%
  \BibitemOpen
  \bibfield{author}{%
  \bibinfo {author} {\bibfnamefont{Z.~Y.}\ \bibnamefont{Xie}}, \bibinfo
  {author} {\bibfnamefont{H.~C.}\ \bibnamefont{Jiang}}, \bibinfo {author}
  {\bibfnamefont{Q.~N.}\ \bibnamefont{Chen}}, \bibinfo {author}
  {\bibfnamefont{Z.~Y.}\ \bibnamefont{Weng}},\ and\ \bibinfo {author}
  {\bibfnamefont{T.}~\bibnamefont{Xiang}},\ }%
  \bibfield{journal}{%
  \bibinfo {journal} {Phys. Rev. Lett.}\ }%
  \textbf{\bibinfo {volume} {103}},\ \bibinfo {pages} {160601} (\bibinfo {year}
  {2009})%
  \bibAnnoteFile{NoStop}{xie2009}%
\bibitem{jordan2009}%
  \BibitemOpen
  \bibfield{author}{%
  \bibinfo {author} {\bibfnamefont{J.}~\bibnamefont{Jordan}}, \bibinfo {author}
  {\bibfnamefont{R.}~\bibnamefont{Or\'us}},\ and\ \bibinfo {author}
  {\bibfnamefont{G.}~\bibnamefont{Vidal}},\ }%
  \bibfield{journal}{%
  \bibinfo {journal} {Phys. Rev. B}\ }%
  \textbf{\bibinfo {volume} {79}},\ \bibinfo {pages} {174515} (\bibinfo {year}
  {2009})%
  \bibAnnoteFile{NoStop}{jordan2009}%
\bibitem{orus2009}%
  \BibitemOpen
  \bibfield{author}{%
  \bibinfo {author} {\bibfnamefont{R.}~\bibnamefont{Or\'{u}s}}, \bibinfo
  {author} {\bibfnamefont{A.~C.}\ \bibnamefont{Doherty}},\ and\ \bibinfo
  {author} {\bibfnamefont{G.}~\bibnamefont{Vidal}},\ }%
  \bibfield{journal}{%
  \bibinfo {journal} {Phys. Rev. Lett.}\ }%
  \textbf{\bibinfo {volume} {102}},\ \bibinfo {pages} {077203} (\bibinfo {year}
  {2009})%
  \bibAnnoteFile{NoStop}{orus2009}%
\bibitem{orus2009-1}%
  \BibitemOpen
  \bibfield{author}{%
  \bibinfo {author} {\bibfnamefont{R.}~\bibnamefont{Or\'us}}\ and\ \bibinfo
  {author} {\bibfnamefont{G.}~\bibnamefont{Vidal}},\ }%
  \bibfield{journal}{%
  \bibinfo {journal} {Phys. Rev. B}\ }%
  \textbf{\bibinfo {volume} {80}},\ \bibinfo {pages} {094403} (\bibinfo {year}
  {2009})%
  \bibAnnoteFile{NoStop}{orus2009-1}%
\bibitem{bauer2009}%
  \BibitemOpen
  \bibfield{author}{%
  \bibinfo {author} {\bibfnamefont{B.}~\bibnamefont{Bauer}}, \bibinfo {author}
  {\bibfnamefont{G.}~\bibnamefont{Vidal}},\ and\ \bibinfo {author}
  {\bibfnamefont{M.}~\bibnamefont{Troyer}},\ }%
  \bibfield{journal}{%
  \bibinfo {journal} {J. Stat. Mech.},\ \bibinfo {pages} {P09006}}%
   (\bibinfo {year} {2009})%
  \bibAnnoteFile{NoStop}{bauer2009}%
\bibitem{pchen2010}%
  \BibitemOpen
  \bibfield{author}{%
  \bibinfo {author} {\bibfnamefont{P.}~\bibnamefont{Chen}}, \bibinfo {author}
  {\bibfnamefont{C.-Y.}\ \bibnamefont{Lai}},\ and\ \bibinfo {author}
  {\bibfnamefont{M.-F.}\ \bibnamefont{Yang}},\ }%
  \bibfield{journal}{%
  \bibinfo {journal} {Phys. Rev. B}\ }%
  \textbf{\bibinfo {volume} {81}},\ \bibinfo {pages} {020409} (\bibinfo {year}
  {2010})%
  \bibAnnoteFile{NoStop}{pchen2010}%
\bibitem{vidal2007-1}%
  \BibitemOpen
  \bibfield{author}{%
  \bibinfo {author} {\bibfnamefont{G.}~\bibnamefont{Vidal}},\ }%
  \bibfield{journal}{%
  \bibinfo {journal} {Phys. Rev. Lett.}\ }%
  \textbf{\bibinfo {volume} {99}},\ \bibinfo {pages} {220405} (\bibinfo {year}
  {2007})%
  \bibAnnoteFile{NoStop}{vidal2007-1}%
\bibitem{vidal2008}%
  \BibitemOpen
  \bibfield{author}{%
  \bibinfo {author} {\bibfnamefont{G.}~\bibnamefont{Vidal}},\ }%
  \bibfield{journal}{%
  \bibinfo {journal} {Phys. Rev. Lett.}\ }%
  \textbf{\bibinfo {volume} {101}},\ \bibinfo {pages} {110501} (\bibinfo {year}
  {2008})%
  \bibAnnoteFile{NoStop}{vidal2008}%
\bibitem{evenbly2009-1}%
  \BibitemOpen
  \bibfield{author}{%
  \bibinfo {author} {\bibfnamefont{G.}~\bibnamefont{Evenbly}}\ and\ \bibinfo
  {author} {\bibfnamefont{G.}~\bibnamefont{Vidal}},\ }%
  \bibfield{journal}{%
  \bibinfo {journal} {Phys. Rev. B}\ }%
  \textbf{\bibinfo {volume} {79}},\ \bibinfo {pages} {144108} (\bibinfo {year}
  {2009})%
  \bibAnnoteFile{NoStop}{evenbly2009-1}%
\bibitem{evenbly2009-2}%
  \BibitemOpen
  \bibfield{author}{%
  \bibinfo {author} {\bibfnamefont{G.}~\bibnamefont{Evenbly}}\ and\ \bibinfo
  {author} {\bibfnamefont{G.}~\bibnamefont{Vidal}},\ }%
  \bibfield{journal}{%
  \bibinfo {journal} {Phys. Rev. Lett.}\ }%
  \textbf{\bibinfo {volume} {102}},\ \bibinfo {pages} {180406} (\bibinfo {year}
  {2009})%
  \bibAnnoteFile{NoStop}{evenbly2009-2}%
\bibitem{evenbly2010}%
  \BibitemOpen
  \bibfield{author}{%
  \bibinfo {author} {\bibfnamefont{G.}~\bibnamefont{Evenbly}}\ and\ \bibinfo
  {author} {\bibfnamefont{G.}~\bibnamefont{Vidal}},\ }%
  \bibfield{journal}{%
  \bibinfo {journal} {Phys. Rev. Lett.}\ }%
  \textbf{\bibinfo {volume} {104}},\ \bibinfo {pages} {187203} (\bibinfo {year}
  {2010})%
  \bibAnnoteFile{NoStop}{evenbly2010}%
\bibitem{pfeifer2009}%
  \BibitemOpen
  \bibfield{author}{%
  \bibinfo {author} {\bibfnamefont{R.~N.~C.}\ \bibnamefont{Pfeifer}}, \bibinfo
  {author} {\bibfnamefont{G.}~\bibnamefont{Evenbly}},\ and\ \bibinfo {author}
  {\bibfnamefont{G.}~\bibnamefont{Vidal}},\ }%
  \bibfield{journal}{%
  \bibinfo {journal} {Phys. Rev. A}\ }%
  \textbf{\bibinfo {volume} {79}},\ \bibinfo {pages} {040301} (\bibinfo {year}
  {2009})%
  \bibAnnoteFile{NoStop}{pfeifer2009}%
\bibitem{schollwoeck2005}%
  \BibitemOpen
  \bibfield{author}{%
  \bibinfo {author} {\bibfnamefont{U.}~\bibnamefont{Schollw{\"o}ck}},\ }%
  \bibfield{journal}{%
  \bibinfo {journal} {Rev. Mod. Phys.}\ }%
  \textbf{\bibinfo {volume} {77}},\ \bibinfo {pages} {259} (\bibinfo {year}
  {2005})%
  \bibAnnoteFile{NoStop}{schollwoeck2005}%
\bibitem{sierra1997}%
  \BibitemOpen
  \bibfield{author}{%
  \bibinfo {author} {\bibfnamefont{G.}~\bibnamefont{Sierra}}\ and\ \bibinfo
  {author} {\bibfnamefont{T.}~\bibnamefont{Nishino}},\ }%
  \bibfield{journal}{%
  \bibinfo {journal} {Nuclear Physics B}\ }%
  \textbf{\bibinfo {volume} {495}},\ \bibinfo {pages} {505 } (\bibinfo {year}
  {1997})%
  \bibAnnoteFile{NoStop}{sierra1997}%
\bibitem{dukelsky1998}%
  \BibitemOpen
  \bibfield{author}{%
  \bibinfo {author} {\bibfnamefont{J.}~\bibnamefont{Dukelsky}}, \bibinfo
  {author} {\bibfnamefont{M.~A.}\ \bibnamefont{Mart{\'\i}n-Delgado}}, \bibinfo
  {author} {\bibfnamefont{T.}~\bibnamefont{Nishino}},\ and\ \bibinfo {author}
  {\bibfnamefont{G.}~\bibnamefont{Sierra}},\ }%
  \bibfield{journal}{%
  \bibinfo {journal} {EPL (Europhysics Letters)}\ }%
  \textbf{\bibinfo {volume} {43}},\ \bibinfo {pages} {457} (\bibinfo {year}
  {1998})%
  \bibAnnoteFile{NoStop}{dukelsky1998}%
\bibitem{wada2000}%
  \BibitemOpen
  \bibfield{author}{%
  \bibinfo {author} {\bibfnamefont{W.}~\bibnamefont{Tatsuaki}},\ }%
  \bibfield{journal}{%
  \bibinfo {journal} {Phys. Rev. E}\ }%
  \textbf{\bibinfo {volume} {61}},\ \bibinfo {pages} {3199} (\bibinfo {year}
  {2000})%
  \bibAnnoteFile{NoStop}{wada2000}%
\bibitem{wada2001}%
  \BibitemOpen
  \bibfield{author}{%
  \bibinfo {author} {\bibfnamefont{W.}~\bibnamefont{Tatsuaki}}\ and\ \bibinfo
  {author} {\bibfnamefont{T.}~\bibnamefont{Nishino}},\ }%
  \bibfield{journal}{%
  \bibinfo {journal} {Computer Physics Communications}\ }%
  \textbf{\bibinfo {volume} {142}},\ \bibinfo {pages} {164 } (\bibinfo {year}
  {2001})%
  \bibAnnoteFile{NoStop}{wada2001}%
\bibitem{mcculloch2000}%
  \BibitemOpen
  \bibfield{author}{%
  \bibinfo {author} {\bibfnamefont{I.~P.}\ \bibnamefont{McCulloch}}\ and\
  \bibinfo {author} {\bibfnamefont{M.}~\bibnamefont{Gul\'acsi}},\ }%
  \bibfield{journal}{%
  \bibinfo {journal} {Aust. J. Phys.}\ }%
  \textbf{\bibinfo {volume} {53}},\ \bibinfo {pages} {597} (\bibinfo {year}
  {2000})%
  \bibAnnoteFile{NoStop}{mcculloch2000}%
\bibitem{mcculloch2001}%
  \BibitemOpen
  \bibfield{author}{%
  \bibinfo {author} {\bibfnamefont{I.~P.}\ \bibnamefont{McCulloch}}\ and\
  \bibinfo {author} {\bibfnamefont{M.}~\bibnamefont{Gul\'acsi}},\ }%
  \bibfield{journal}{%
  \bibinfo {journal} {Phil. Mag. Lett.}\ }%
  \textbf{\bibinfo {volume} {81}},\ \bibinfo {pages} {447 } (\bibinfo {year}
  {2001})%
  \bibAnnoteFile{NoStop}{mcculloch2001}%
\bibitem{mcculloch2002}%
  \BibitemOpen
  \bibfield{author}{%
  \bibinfo {author} {\bibfnamefont{I.~P.}\ \bibnamefont{McCulloch}}\ and\
  \bibinfo {author} {\bibfnamefont{M.}~\bibnamefont{Gul{\'a}csi}},\ }%
  \bibfield{journal}{%
  \bibinfo {journal} {EPL (Europhysics Letters)}\ }%
  \textbf{\bibinfo {volume} {57}},\ \bibinfo {pages} {852} (\bibinfo {year}
  {2002})%
  \bibAnnoteFile{NoStop}{mcculloch2002}%
\bibitem{kraus2010}%
  \BibitemOpen
  \bibfield{author}{%
  \bibinfo {author} {\bibfnamefont{C.~V.}\ \bibnamefont{Kraus}}, \bibinfo
  {author} {\bibfnamefont{N.}~\bibnamefont{Schuch}}, \bibinfo {author}
  {\bibfnamefont{F.}~\bibnamefont{Verstraete}},\ and\ \bibinfo {author}
  {\bibfnamefont{J.~I.}\ \bibnamefont{Cirac}},\ }%
  \bibfield{journal}{%
  \bibinfo {journal} {Phys. Rev. A}\ }%
  \textbf{\bibinfo {volume} {81}},\ \bibinfo {pages} {052338} (\bibinfo {year}
  {2010})%
  \bibAnnoteFile{NoStop}{kraus2010}%
\bibitem{barthel2009}%
  \BibitemOpen
  \bibfield{author}{%
  \bibinfo {author} {\bibfnamefont{T.}~\bibnamefont{Barthel}}, \bibinfo
  {author} {\bibfnamefont{C.}~\bibnamefont{Pineda}},\ and\ \bibinfo {author}
  {\bibfnamefont{J.}~\bibnamefont{Eisert}},\ }%
  \bibfield{journal}{%
  \bibinfo {journal} {Phys. Rev. A}\ }%
  \textbf{\bibinfo {volume} {80}},\ \bibinfo {pages} {042333} (\bibinfo {year}
  {2009})%
  \bibAnnoteFile{NoStop}{barthel2009}%
\bibitem{corboz2009}%
  \BibitemOpen
  \bibfield{author}{%
  \bibinfo {author} {\bibfnamefont{P.}~\bibnamefont{Corboz}}, \bibinfo {author}
  {\bibfnamefont{G.}~\bibnamefont{Evenbly}}, \bibinfo {author}
  {\bibfnamefont{F.}~\bibnamefont{Verstraete}},\ and\ \bibinfo {author}
  {\bibfnamefont{G.}~\bibnamefont{Vidal}},\ }%
  \bibfield{journal}{%
  \bibinfo {journal} {Phys. Rev. A}\ }%
  \textbf{\bibinfo {volume} {81}},\ \bibinfo {pages} {010303(R)} (\bibinfo
  {year} {2010})%
  \bibAnnoteFile{NoStop}{corboz2009}%
\bibitem{corboz2009-1}%
  \BibitemOpen
  \bibfield{author}{%
  \bibinfo {author} {\bibfnamefont{P.}~\bibnamefont{Corboz}}\ and\ \bibinfo
  {author} {\bibfnamefont{G.}~\bibnamefont{Vidal}},\ }%
  \bibfield{journal}{%
  \bibinfo {journal} {Phys. Rev. B}\ }%
  \textbf{\bibinfo {volume} {80}},\ \bibinfo {pages} {165129} (\bibinfo {year}
  {2009})%
  \bibAnnoteFile{NoStop}{corboz2009-1}%
\bibitem{corboz2010}%
  \BibitemOpen
  \bibfield{author}{%
  \bibinfo {author} {\bibfnamefont{P.}~\bibnamefont{Corboz}}, \bibinfo {author}
  {\bibfnamefont{R.}~\bibnamefont{Or\'us}}, \bibinfo {author}
  {\bibfnamefont{B.}~\bibnamefont{Bauer}},\ and\ \bibinfo {author}
  {\bibfnamefont{G.}~\bibnamefont{Vidal}},\ }%
  \bibfield{journal}{%
  \bibinfo {journal} {Phys. Rev. B}\ }%
  \textbf{\bibinfo {volume} {81}},\ \bibinfo {pages} {165104} (\bibinfo {year}
  {2010})%
  \bibAnnoteFile{NoStop}{corboz2010}%
\bibitem{shi2009}%
  \BibitemOpen
  \bibfield{author}{%
  \bibinfo {author} {\bibfnamefont{Q.-Q.}\ \bibnamefont{Shi}}, \bibinfo
  {author} {\bibfnamefont{S.-H.}\ \bibnamefont{Li}}, \bibinfo {author}
  {\bibfnamefont{J.-H.}\ \bibnamefont{Zhao}},\ and\ \bibinfo {author}
  {\bibfnamefont{H.-Q.}\ \bibnamefont{Zhou}},\ }%
  \bibfield{journal}{%
  \bibinfo {journal} {Preprint}}%
   (\bibinfo {year} {2009}),\ \bibinfo {note} {arXiv:0907.5520}%
  \bibAnnoteFile{NoStop}{shi2009}%
\bibitem{pizorn2010}%
  \BibitemOpen
  \bibfield{author}{%
  \bibinfo {author} {\bibfnamefont{I.}~\bibnamefont{Pi\v{z}orn}}\ and\ \bibinfo
  {author} {\bibfnamefont{F.}~\bibnamefont{Verstraete}},\ }%
  \bibfield{journal}{%
  \bibinfo {journal} {Phys. Rev. B}\ }%
  \textbf{\bibinfo {volume} {81}},\ \bibinfo {pages} {245110} (\bibinfo {year}
  {2010})%
  \bibAnnoteFile{NoStop}{pizorn2010}%
\bibitem{gu2010}%
  \BibitemOpen
  \bibfield{author}{%
  \bibinfo {author} {\bibfnamefont{Z.-C.}\ \bibnamefont{Gu}}, \bibinfo {author}
  {\bibfnamefont{F.}~\bibnamefont{Verstraete}},\ and\ \bibinfo {author}
  {\bibfnamefont{X.-G.}\ \bibnamefont{Wen}},\ }%
  \bibfield{journal}{%
  \bibinfo {journal} {Preprint}}%
   (\bibinfo {year} {2010}),\ \bibinfo {note} {arXiv:1004.2563}%
  \bibAnnoteFile{NoStop}{gu2010}%
\bibitem{cincio2008}%
  \BibitemOpen
  \bibfield{author}{%
  \bibinfo {author} {\bibfnamefont{L.}~\bibnamefont{Cincio}}, \bibinfo {author}
  {\bibfnamefont{J.}~\bibnamefont{Dziarmaga}},\ and\ \bibinfo {author}
  {\bibfnamefont{M.~M.}\ \bibnamefont{Rams}},\ }%
  \bibfield{journal}{%
  \bibinfo {journal} {Phys. Rev. Lett.}\ }%
  \textbf{\bibinfo {volume} {100}},\ \bibinfo {pages} {240603} (\bibinfo
  {month} {Jun}\ \bibinfo {year} {2008})%
  \bibAnnoteFile{NoStop}{cincio2008}%
\bibitem{zhao2010}%
  \BibitemOpen
  \bibfield{author}{%
  \bibinfo {author} {\bibfnamefont{H.~H.}\ \bibnamefont{Zhao}}, \bibinfo
  {author} {\bibfnamefont{Z.~Y.}\ \bibnamefont{Xie}}, \bibinfo {author}
  {\bibfnamefont{Q.~N.}\ \bibnamefont{Chen}}, \bibinfo {author}
  {\bibfnamefont{Z.~C.}\ \bibnamefont{Wei}}, \bibinfo {author}
  {\bibfnamefont{J.~W.}\ \bibnamefont{Cai}},\ and\ \bibinfo {author}
  {\bibfnamefont{T.}~\bibnamefont{Xiang}},\ }%
  \bibfield{journal}{%
  \bibinfo {journal} {Phys. Rev. B}\ }%
  \textbf{\bibinfo {volume} {81}},\ \bibinfo {pages} {174411} (\bibinfo {year}
  {2010})%
  \bibAnnoteFile{NoStop}{zhao2010}%
\bibitem{evenbly2010-1}%
  \BibitemOpen
  \bibfield{author}{%
  \bibinfo {author} {\bibfnamefont{G.}~\bibnamefont{Evenbly}}, \bibinfo
  {author} {\bibfnamefont{P.}~\bibnamefont{Corboz}},\ and\ \bibinfo {author}
  {\bibfnamefont{G.}~\bibnamefont{Vidal}},\ }%
  \bibfield{journal}{%
  \bibinfo {journal} {Phys. Rev. B}\ }%
  \textbf{\bibinfo {volume} {82}},\ \bibinfo {pages} {132411} (\bibinfo {year}
  {2010})%
  \bibAnnoteFile{NoStop}{evenbly2010-1}%
\bibitem{singh2010}%
  \BibitemOpen
  \bibfield{author}{%
  \bibinfo {author} {\bibfnamefont{S.}~\bibnamefont{Singh}}, \bibinfo {author}
  {\bibfnamefont{R.~N.~C.}\ \bibnamefont{Pfeifer}},\ and\ \bibinfo {author}
  {\bibfnamefont{G.}~\bibnamefont{Vidal}},\ }%
  \bibfield{journal}{%
  \bibinfo {journal} {Preprint}}%
   (\bibinfo {year} {2010}),\ \bibinfo {note} {arXiv:1008.4774}%
  \bibAnnoteFile{NoStop}{singh2010}%
\bibitem{singh2010-1}%
  \BibitemOpen
  \bibfield{author}{%
  \bibinfo {author} {\bibfnamefont{S.}~\bibnamefont{Singh}}, \bibinfo {author}
  {\bibfnamefont{R.~N.~C.}\ \bibnamefont{Pfeifer}},\ and\ \bibinfo {author}
  {\bibfnamefont{G.}~\bibnamefont{Vidal}},\ }%
  \bibfield{journal}{%
  \bibinfo {journal} {Phys. Rev. A}\ }%
  \textbf{\bibinfo {volume} {82}},\ \bibinfo {pages} {050301} (\bibinfo {year}
  {2010})%
  \bibAnnoteFile{NoStop}{singh2010-1}%
\bibitem{jordan2008}%
  \BibitemOpen
  \bibfield{author}{%
  \bibinfo {author} {\bibfnamefont{J.}~\bibnamefont{Jordan}}, \bibinfo {author}
  {\bibfnamefont{R.}~\bibnamefont{Or\'{u}s}}, \bibinfo {author}
  {\bibfnamefont{G.}~\bibnamefont{Vidal}}, \bibinfo {author}
  {\bibfnamefont{F.}~\bibnamefont{Verstraete}},\ and\ \bibinfo {author}
  {\bibfnamefont{J.~I.}\ \bibnamefont{Cirac}},\ }%
  \bibfield{journal}{%
  \bibinfo {journal} {Phys. Rev. Lett.}\ }%
  \textbf{\bibinfo {volume} {101}},\ \bibinfo {pages} {250602} (\bibinfo {year}
  {2008})%
  \bibAnnoteFile{NoStop}{jordan2008}%
\bibitem{sandvik1997}%
  \BibitemOpen
  \bibfield{author}{%
  \bibinfo {author} {\bibfnamefont{A.~W.}\ \bibnamefont{Sandvik}},\ }%
  \bibfield{journal}{%
  \bibinfo {journal} {Phys. Rev. B}\ }%
  \textbf{\bibinfo {volume} {56}},\ \bibinfo {pages} {11678} (\bibinfo {year}
  {1997})%
  \bibAnnoteFile{NoStop}{sandvik1997}%
\bibitem{sandvik1999}%
  \BibitemOpen
  \bibfield{author}{%
  \bibinfo {author} {\bibfnamefont{A.~W.}\ \bibnamefont{Sandvik}}\ and\
  \bibinfo {author} {\bibfnamefont{C.~J.}\ \bibnamefont{Hamer}},\ }%
  \bibfield{journal}{%
  \bibinfo {journal} {Phys. Rev. B}\ }%
  \textbf{\bibinfo {volume} {60}},\ \bibinfo {pages} {6588} (\bibinfo {year}
  {1999})%
  \bibAnnoteFile{NoStop}{sandvik1999}%
\end{thebibliography}%
\bibliographystyle{apsrev4-1}

\end{document}